\documentclass[
aps,
prb,
 twocolumn,
 showpacs,
 superscriptaddress,
]{revtex4-1}

\usepackage[latin1]{inputenc}

\usepackage[ english ]{babel}

\usepackage{ragged2e}

\usepackage[ final ]{graphicx}

\usepackage[ centertags, sumlimits, intlimits, namelimits, fleqn, ]{amsmath}

\usepackage{amssymb}

\usepackage{dcolumn}

\usepackage{natbib}

\begin{document} \title{A quantized current source with mesoscopic feedback}
\date{\today}

\author{Lukas \surname{Fricke}}\email[Present address: Physikalisch-Technische
Bundesanstalt, Bundesallee 100, 38116 Braunschweig, Germany. Mail to:
]{lukas.fricke@ptb.de}\affiliation{Institut f\"ur Festk\"orperphysik, Leibniz
Universit\"at Hannover, 30167 Hannover, Germany.}\noaffiliation

\author{Frank \surname{Hohls}}\affiliation{Institut f\"ur Festk\"orperphysik,
Leibniz Universit\"at Hannover, 30167 Hannover,
Germany.}\affiliation{Physikalisch-Technische Bundesanstalt, Bundesallee 100,
38116 Braunschweig, Germany.}\noaffiliation

\author{Niels \surname{Ubbelohde}}\affiliation{Institut f\"ur
Festk\"orperphysik, Leibniz Universit\"at Hannover, 30167 Hannover, Germany.}\noaffiliation

\author{Bernd \surname{Kaestner}}\affiliation{Physikalisch-Technische
Bundesanstalt, Bundesallee 100, 38116 Braunschweig, Germany.}\noaffiliation

\author{Vyacheslavs \surname{Kashcheyevs}}\affiliation{Faculty of Physics and Mathematics,
University of Latvia, Riga LV-1586, Latvia.}\noaffiliation

\author{Christoph \surname{Leicht}}\affiliation{Physikalisch-Technische
Bundesanstalt, Bundesallee 100, 38116 Braunschweig, Germany.}\noaffiliation

\author{Philipp \surname{Mirovsky}}\affiliation{Physikalisch-Technische
Bundesanstalt, Bundesallee 100, 38116 Braunschweig, Germany.}\noaffiliation

\author{Klaus \surname{Pierz}}\affiliation{Physikalisch-Technische
Bundesanstalt, Bundesallee 100, 38116 Braunschweig, Germany.}\noaffiliation

\author{Hans W. \surname{Schumacher}}\affiliation{Physikalisch-Technische
Bundesanstalt, Bundesallee 100, 38116 Braunschweig, Germany.}\noaffiliation

\author{Rolf J. \surname{Haug}}\affiliation{Institut f\"ur Festk\"orperphysik,
Leibniz Universit\"at Hannover, 30167 Hannover, Germany.}\noaffiliation

%\pacs{73.23.-b, 72.10.-d, 73.22.Dj, 73.63.Kv}

%73.23.-b Electronic transport in mesoscopic systems

%72.10.-d Theory of electronic transport; scattering mechanisms

%73.22.Dj Electronic structure of nanoscale materials and related systems : Single particle states

%73.63.Kv Electronic transport in nanoscale materials and structures : Quantum dots

%%%%%%%%%%% Abstract %%%%%%%%%%%%%%%%%%%%%%%%%%%%%%%%%
\begin{abstract} We study a mesoscopic circuit of two quantized current sources,
realized by non-adiabatic single-electron pumps connected in series with a
small micron-sized island in between. We find that quantum transport through
the second pump can be locked onto the quantized current of the first one by a
feedback due to charging of the mesoscopic island. This is confirmed by a
measurement of the charge variation on the island using a nearby charge
detector. Finally, the charge feedback signal clearly evidences loading into
excited states of the dynamic quantum dot during single-electron pump
operation.
\end{abstract}\maketitle

%\section{Introduction}
%%%%%%%%%%%%%%%%%%% Introduction %%%%%%%%%%%%%%%%%%%%%%%%%%%
Quantized current sources (e.g., see
Refs.~\onlinecite{Pothier92,Blumenthal2007,Kaestner2008PRB,fujiwara2008}) have
interesting applications e.g.\ as on demand electron source for quantum
information processing~\cite{Barnes2000} or as current source for
metrology.~\cite{Mills2006} In contrast to
turnstiles~\cite{Geerligs90,Devoret1992} the current can be driven against a
voltage applied across the pump, allowing their use as current source in
mesoscopic quantum electronics. Yet the further suppression of still present
current fluctuations, as desired for such applications, remains challenging.
Now Brandes has proposed a new method to stabilize quantum transport in
mesoscopic devices against fluctuations, named mesoscopic
feedback.~\cite{Brandes2010} In this paper, we experimentally realize and investigate a quantum transport device with mesoscopic feedback. The device under study is a semiconductor non-adiabatic quantized charge pump. A mesoscopic feedback loop is realized using two quantized charge pumps P1, P2 connected in series and separated by a mesoscopic island in between. In this circuit, any momentary difference between the currents through the two charge pumps immediately leads to a charge accumulation on the mesoscopic island. This in turn acts as feedback mainly onto pump P2, locking it to the nominal current set by pump P1. The charge on the mesoscopic island is monitored by a nearby capacitively coupled detector allowing us to verify the feedback mechanism. Furthermore, the highly sensitive feedback signal reveals a fine structure within the quantized current plateau due to loading into excited states during the initial phase of the pumping cycle, which is not observable in measurements of the pumped current. This demonstrates the use of mesoscopic feedback control as a characterization tool for dynamic processes in nanostructures and, furthermore, it opens the possibility of non-ground-state initialization of dynamic quantum dots with possible applications in quantum information processing. Moreover, this realization of a mesoscopic circuit of quantized charge pumps demonstrates their usability for future applications in integrated mesoscopic electronics.

%\section{Experimental}
%%%%%%%%%%%%%%%%%%%  Device and basics %%%%%%%%%%%%%%%%%%%%%%%%
\begin{figure}[b!] \centering
\centering \includegraphics[width=0.9\linewidth]{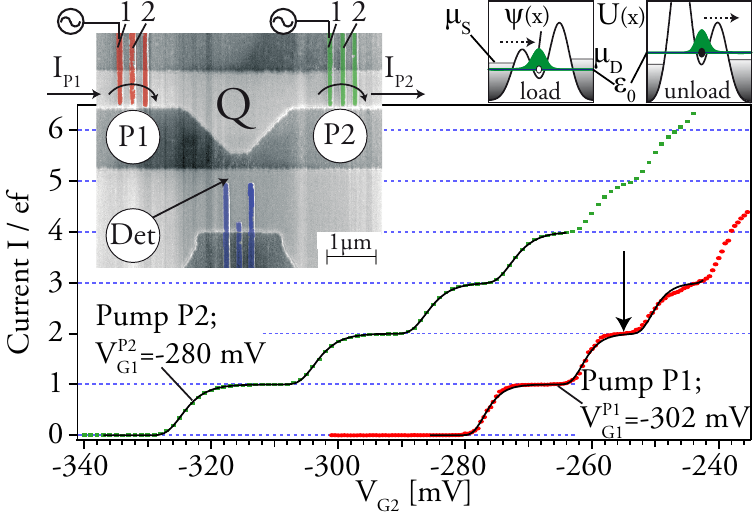} \caption{
(Color online)
Left inset: SEM image. Bright regions are
semiconducting channels. Schottky gates (vertical lines, colored) are used to
realize the single electron pumps (P1, P2) in the upper channel and the detector
(Det) in the lower one.
Right inset: Sketch of the loading and unloading process.
Main figure: DC-Gate characteristic of two single non-adiabatic quantized-charge pumps
(red dots: P1 with \mbox{$V^\text{P1}_\text{G1}=-302$~mV}; green squares:
P2 with \mbox{$V^\text{P2}_\text{G1}=-280$~mV}; solid lines: model fits). The
black arrow indicates the P1 working point for series operation.
\label{SEM} \label{single}} \end{figure}

The device is shown in Fig.~\ref{single}: Two dynamically driven quantum
dots, each formed by titanium Schottky gates (colored lines) across a narrow
semiconducting channel, act as single-electron pumps. They are connected by a
few micron wide mesoscopic island that is capacitively coupled to a charge
detector, also realized by a gated channel.  The channels were etched from a
two-dimensional electron system in an Al$_{0.3}$Ga$_{0.7}$As/GaAs
heterostructure 90~nm below the surface and connected individually via wide
leads to ohmic contacts. The device was measured at a temperature of 1.5 K
immersed in liquid helium and the frequency and amplitude of the oscillating
pump voltage were always kept at $f=50$~MHz respective $V_{ac} \approx 50$~mV. When both pumps are operating, they are driven by the same ac source.

We will first recapitulate the operation of the individual
pump~\cite{Kaestner2008PRB,kashcheyevs2009}: Negative voltages are applied to
two Schottky gates (labeled 1 and 2) forming a quantum dot (QD) in the channel,
the third gate is grounded. A sinusoidal signal with frequency $f$ is
superimposed onto gate 1 (G1) modulating both the height of that barrier and
the QD potential. In the loading phase during the first half-cycle (sketch in
Fig.~\ref{single}) this modulation drives the QD energy $\varepsilon_0$ below
the source chemical potential $\mu_s$ and electrons are loaded onto the dot.
Raising the barrier also raises the dot potential. A cascade of electron
back-tunnelling to source sets in when $\varepsilon_0$ crosses $\mu_s$. Further
rising of the barrier strongly reduces the decay rate and the cascade stops at
$n$ captured electrons, where $n$ can be controlled by the voltage on gate 2,
$V_\text{G2}$~\cite{kashcheyevs2009}. The trapped electrons are subsequently
ejected into drain, generating a quantized current $I=nef$ with $e$ the
elementary charge. The staircase-like current dependence on $V_\text{G2}$ is
shown in Fig.~\ref{single} for the individual pumps together with a fit to
$I(V_\text{G2})=ef\sum_{l=0}^{N}\exp(-\exp[-\alpha
V_\text{G2}+\sum_{k=0}^l\delta_k])$ as derived in Ref.~\onlinecite{kashcheyevs2009} with $N$ determined by the number of plateaus to be fitted and fitting parameters $\alpha$ and $\delta_k$. The double-exponential function relates the electron current via decay rates during the initialization process (isolation of $n$ electrons on the dot) to the QD energy levels controlled by the applied gate voltage $V_\text{G2}$ scaled with $\alpha$. The fitting parameters $\delta_k$ correspond to the positions of the plateau transitions.

For a moderate change of the source or drain potential by $\Delta\mu_{s/d}$
only the positions of the current steps are shifted,
$I(V_\text{G2},\Delta\mu_s,\Delta\mu_d) =
I(V_\text{G2}+\beta_s\Delta\mu_s-\beta_d\Delta\mu_d)$ with the lever arms
$\beta_{s/d}$ describing the coupling between a potential $-\Delta\mu/e$ on the source and drain, respectively, and the dynamic dot. Additionally, we expect a stronger effect of the source chemical potential (loading side) onto the decay cascade phase with opposite sign compared to gating: a rise of the potential,
$\Delta\mu_s > 0$, leads to a delayed start of the decay cascade phase and
therefore lower back-tunneling rates for initially captured electrons. Thus the
average number of electrons kept till unloading is increased, yielding a larger
current~\cite{kashcheyevs2009}. The argument for $\Delta\mu_s < 0$ is
analogues. No similar effect is expected for the unloading side as electrons
are ejected with rather high excess energies~\cite{Leicht2011}. Thus we expect
$\beta_s \gg \beta_d>0$.

%%%%%%%%%%%%%%%%% Series Pumping %%%%%%%%%%%%%%%%%%%%%
%\section{Results and Discussion}
\begin{figure}[t!b] \centering \includegraphics[]{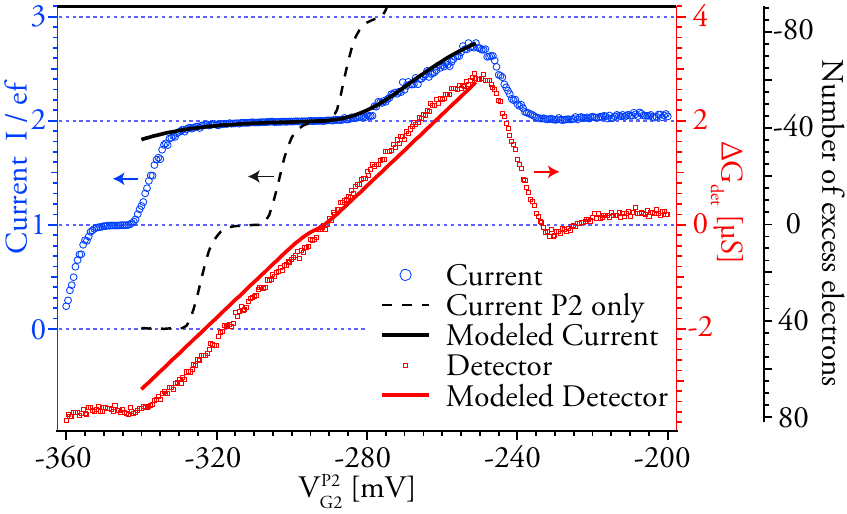}
\caption{ (Color online) Two single-electron pumps in series with P1 set to a
working point with $I=2ef$ in single operation as marked in Fig. \ref{single}.
Comparison of serial and single pump operation of P2 with
$V^\text{P2}_\text{G1}=-280$~mV (circles resp.\ short dashed line, left axis)
and the corresponding charge detector signal in serial operation (small squares, right axis). Solid lines show modeled current and charge signal as discussed in the text. The second (rightmost) scale for the detector signal
shows the number of excess electrons on the island using $\Delta G/\Delta\mu=0.26\mu\text{S}/\text{meV}$ and $C=0.9\pm0.1\;$fF. The present resolution limit due to detector noise is $\Delta N\approx 2$. \label{ser1}} \end{figure}

We now turn to the series operation with pump P1 set to a fixed working point
with a nominal current of $I=2ef$ while varying $V_{G2}$ of pump P2.
Figure~\ref{ser1} shows the resulting gate-dependence of the current through
both pumps connected in series (circles) in comparison to the single pump P2
(dashed line). One immediately notices a largely (fivefold) extended $2ef$
plateau in the $V^\text{P2}_\text{G2}$ dependence. In the following we will
explain this due to the asymmetric feedback effect of charge accumulated on the
mesoscopic island in between the pumps.

An electron current $I_\text{P1}$ is pumped by P1 on the island and $I_\text{P2}$ is pumped by P2
off the island. Any momentary difference $\delta I(t) = I_\text{P2}(t) -
I_\text{P1}(t)$ between the currents leads to a charge built up $Q = -\int
\delta I(t)\,dt$ with accompanying change $\Delta\mu = -eQ/C$ of the potential
on the mesoscopic island with total capacitance $C$. This leads to an effective
bias with opposite sign across the two pumps, acting as feedback onto the
currents produced by P1 and P2. A steady state is reached when the average
currents of both pumps are equal, i.e.
$\overline{I}_\text{P1}(V_\text{G2}^\text{P1}-\beta_d\overline{\Delta\mu}) =
\overline{I}_\text{P2}(V_\text{G2}^\text{P2}+\beta_s\overline{\Delta\mu})$. Due
to $\beta_s \gg \beta_d$ the charging of the island predominantly acts as an
efficient feedback mechanism to control P2, stabilizing it onto the $2ef$
plateau over a wide gate voltage range, while P1 acts as a mesoscopic current
source with very large impedance against the island potential.

For a detailed study of this feedback, we have additionally integrated a charge
detector in our device: A second semiconductor channel close to the island can
be depleted near to pinch off by Schottky gates (see the left inset in Fig~\ref{single}). The potential due to an accumulated charge on the island changes
the conductance $G_\text{det}$ of the detector channel.~\cite{Field1993} This
is measured using a lock-in amplifier with an applied source-drain voltage of
$V_\text{det}=30\ \mu$V at a frequency $f_\text{det}=213$~Hz. We determine the
detector sensitivity using the island with open pump P2 as a side gate. Around
the working point $G_{\det}^0 \approx 50\,\mu$S the detector reacts nearly
linearly to an island potential variation ranging from $-30$~meV to $20$~meV with
a response of $\Delta G/\Delta \mu = -0.26\,\mu\text{S}/\text{meV}$. We have
also determined the crosstalk of the pump gates ($5\cdot
10^{-3}\,\mu\text{S}/\text{mV}$) and subtracted it in the following. For an
estimate of the accumulated charge $Q = -\Delta\mu\cdot C/e$ and the excess
electron number $N=-Q/e$, respectively, we determine the capacitance of the island from a
finite-element solution of the Poisson equation to $C= 0.9\pm0.1$~fF with the
error estimated from the uncertainty in the depletion length at the mesa edge.

The measured charge signal is shown in Fig.~\ref{ser1} by the small squares.
Its zero point is assigned to the intersection of the curves for single and
series pumping ($V^\text{P2}_\text{G2}=-290.5$~mV) where we expect an unbiased
pump P2 and therefore an uncharged island. While the current is nearly constant all over
the extended $2ef$ plateau, the charge respective potential of the
island, measured by the detector, changes almost linearly as
a function of control gate voltage, nicely revealing the above
discussed mechanism of charge feedback in series pumping.

We will now compare the measured gate voltage dependence of both current and
charge to a simple model: The starting point is the analytical representation of
the free running pump currents $I_\text{P1}(V_\text{G2}^\text{P1})$ and
$I_\text{P2}(V_\text{G2}^\text{P2})$ as derived by the fits shown by the lines
in Fig.~\ref{single}. We solve
$\overline{I}_\text{P1}(V_\text{G2}^\text{P1}-\beta_d\overline{\Delta\mu}) =
\overline{I}_\text{P2}(V_\text{G2}^\text{P2}+\beta_s\overline{\Delta\mu})$
numerically with $\beta_d = 0.45\pm0.04$~mV/meV determined experimentally and
$\beta_s$ as the only fitting parameter (the experimental setup did not allow a
source biasing of pump P2). The black solid line in Fig.~\ref{ser1} shows the
modeled current with $\beta_s/\beta_d = 7\pm1$ chosen for best fit to the
current data within the range $-325$~mV~$<V^\text{P2}_\text{G2}<-255$~mV of linear detector response. The large ratio of $\beta_s$ and $\beta_d$ confirms nicely the expected strong asymmetry of the feedback, $\beta_s \gg \beta_d$, acting much stronger on pump P2 than on P1.

We can also transform the modeled dependence of the island's chemical potential, $\overline{\Delta\mu}(V_\text{G2}^\text{P2})$, into a modeled detector response using the experimentally determined response factor $\Delta G/\Delta \mu$. This scaling has an accumulated uncertainty due to the uncertainties in $\Delta G/\Delta \mu$ and the bias lever arm $\beta_d$, both measured with an open island, and in the fitted ratio $\beta_s/\beta_d$. Within this margin we find a good agreement with about 20\% deviation of the slope, confirming our understanding of the feedback mechanism.

Our experiment demonstrates a realization of feedback control of quantum
transport as discussed very recently by Brandes~\cite{Brandes2010}: The current
$I_\text{P2}(t)$ through pump P2 and its fluctuations are governed by the
quantum tunneling between the island and the dynamic quantum
dot.~\cite{Maire2008} This current is locked onto the reference current
$I_\text{P1}(t)$ generated by the first pump, albeit a reference that itself
shows the fluctuations of quantum transport. The momentary current error,
$\delta I(t) = I_\text{P2}(t) - I_\text{P1}(t)$, is integrated to the charge
accumulation $Q(t) = -\int^t_0 \delta I(t')\,dt'$ acting back onto
$I_\text{P2}(t)$, thus an \emph{integral} feedback loop is realized. An
interesting prediction of Brandes is the strong suppression of the fluctuations
of $\delta I(t)$ on long time scales. In our experiment, we have to keep in mind
that fluctuations of our reference current, $I_\text{P1}(t)$, will also induce
fluctuations in $I_\text{P2}(t)$ due to the feedback. To access the noise
suppression in $\delta I(t)$, it is therefore not sufficient to measure only the
fluctuations of the current in one of the leads, as done for a single
pump.~\cite{Maire2008} Instead, one has to either examine both time-dependent
currents or, more conveniently, the charge fluctuations $\delta Q(t)$ measured by
the detector.

We can estimate the effect of feedback on current fluctuations of pump P2 as follows, assuming for simplification an ideal pump P1 with constant $I_\text{P1}=I_0$: Writing $Q(t) =
\overline{Q}+\delta Q(t)$ and linearizing the response of
$I_\text{P2}=I_0+\delta I(t)$ to the charge fluctuations $\delta Q(t)$, we find
$\delta I(t)=\Delta I_\text{P2}(t)+\gamma \delta Q$ with $\Delta
I_\text{P2}(t)$ the intrinsic fluctuations of the free running pump and $\gamma
=e\beta_s
C^{-1}\left.\frac{dI_\text{P2}}{dV_\text{G2}^\text{P2}}\right|_{I_0}$.
Transforming to the frequency domain, we can rewrite $\delta Q = -\delta I/i\omega$
and solve for the current fluctuations $\delta I(\omega) = \Delta
I_\text{P2}(\omega)/(1+\gamma/i\omega)$. Thus the intrinsic fluctuations are
suppressed by a factor $\gamma/\omega$ below a characteristic frequency $\gamma
\sim 10^{5}$~Hz. As the free running pump produces white noise,~\cite{Maire2008}
the fluctuation spectrum in closed feedback should be linear for the current
difference, $\delta I_\text{rms}(\omega)\propto \omega$, and constant (white)
for the charge noise $\delta Q_\text{rms}(\omega)$.

Next, we will briefly discuss the behavior outside the nearly linear detector
response in Fig.~\ref{ser1}: At very negative gate voltages $V^\text{P2}_\text{G2}<-325$~mV, we have
finally accumulated a sufficiently large bias across pump P1 to quench its pump
action. Here we either block the ejection of electrons onto the island or we
reach a threshold of the nonlinear resistance for the backflow of electrons.
Thus the effect of the island potential on pump P1 gets even stronger than on
P2. As a result, we observe current steps down to the $ef$-plateau and to zero
with a characteristic gate dependence nearly unchanged from the single pump action
of P2, accompanied by only small variations of the island potential. At the least
negative voltages $V^\text{P2}_\text{G2} > -250$~mV, the exit barrier of P2 is
lowered too far and a backflow of electrons through P2 onto the island becomes
possible; the current switches back to a quantized value of $2ef$ driven by P1,
accompanied by the almost flat detector signal \mbox{$\Delta G\approx 0$} of a
neutral island.

\begin{figure}[t!b] \centering \includegraphics[]{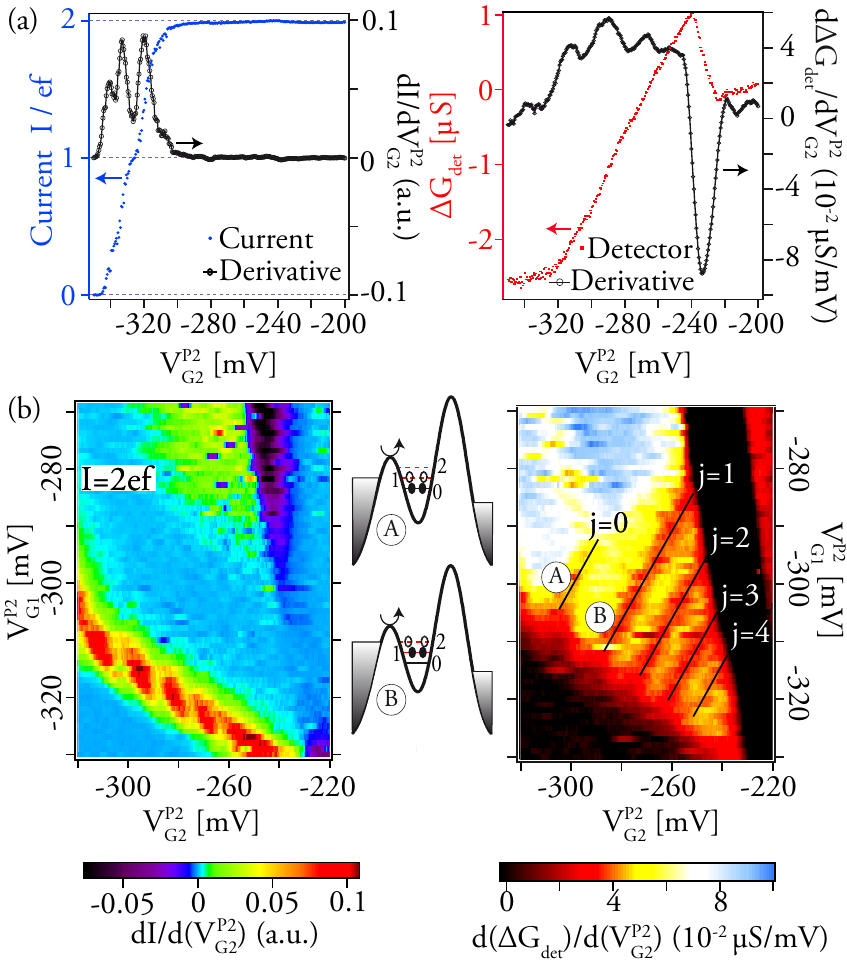}
\caption{(Color online) Two single-electron pumps in series with the same
configuration of pump P1 as in Fig.~\ref{ser1}. (a) Current (left) and detector
signal (right) for \mbox{$V^\text{P2}_\text{G1}=-305$~mV} both together with
derivatives. (b) Derivatives of series pumping current (left) and detector signal (right) as a function of both DC gate voltages of P2. Schematics show loading into ground (A) and first excited state (B) at indicated positions.
\label{ser2}} \end{figure}

Finally, we turn to a fine structure that can be observed in the detector signal
for more negative voltages applied to the entrance gate of P2. Fig.~\ref{ser2}a
shows the current (left) and the charge detector signal (right) and their
derivatives for $V^\text{P2}_\text{G1} = -305$~mV. Focusing on the 2ef current
plateau ($-300\text{ mV}<V^\text{P2}_\text{G2} <-250\text{ mV}$), we observe
minima in the detector derivative $dG_\text{det}/dV^\text{P2}_\text{G2}$ while
the current derivative $dI/dV^\text{P2}_\text{G2}$ hardly shows any structure
on this very flat plateau. The structure can be discerned even better when
varying both gate voltages of pump P2, as shown in Fig.~\ref{ser2}b. Here the
derivative of the detector signal uncovers a set of lines within the 2ef
plateau whereas no structure can be discerned in the current derivative on the
plateau. We note that the lines connect to an oscillating structure at the
plateau edge that here at the edge is also visible in the current derivative. A
line-feature in this plateau region with a positive slope in the
$V_\text{G2}-V_\text{G1}$ parameter space has been predicted from a
single-level model in Ref.~\onlinecite{Leicht2010}. According to this model,
the pumped current should drop to zero below the uppermost line [marked $j=0$ in Fig.~\ref{ser2}(b)] since the source barrier gets too opaque with respect to the ground state during the loading time with \mbox{$\varepsilon_0 < \mu_S$}. Loading of electrons into the ground state is thereby dynamically blockaded.~\cite{Kaestner2008PRB} However, in contradiction to this single level model the current plateau and therefore single electron pumping continues beyond this line. This is due to loading into \emph{excited} states with energy \mbox{$\varepsilon_j>\varepsilon_0$} and higher source barrier transparency. These will allow loading as long as the state $j$ with energy $\varepsilon_j$ can be occupied from source, i.e. it reaches the condition \mbox{$\varepsilon_j < \mu_S$} at some point during the loading process. This condition is always fulfilled at sufficient distance to the plateau onset for some states. The sketches in Fig.~\ref{ser2}(b) illustrate this transition from loading with accessible ground state (A) to a dynamically blockaded ground state but accessible first excited state (B). The effect of loading into excited states overlaps with the termination of loading into the ground state and is therefore unobserved in the direct current measurement. But the resulting change of the loading rate leads to a detectable change of the dynamic charge equilibrium on the island. Thus the charge detection in series setup allows for a new access to the inner structure of the dynamic QD within the quantized current regime. 

Moving further to the lower right in the $V_\text{G2}-V_\text{G1}$ plane also
for the $j=1$ and later for the $j=2$ state and so forth the source barrier
becomes too opaque; the low lying excited states are thus also dynamically
blockaded and loading can only occur into states of successively higher energy,
leading to the additional lines observed. This has a very interesting
implication: Here we initialize excited electron states in the dynamic quantum
dot just by choosing appropriate gate voltages, a first important step toward
the application of these devices for quantum information processing.

%%%%%%%%%%%%%%%%%%% Outlook %%%%%%%%%%%%%%%%%%%%%%%
%\section{Summary}
Coming back to the feedback charge detection we anticipate that in an optimized
geometry and with increased capacitive coupling it will be possible to enhance
the sensitivity of the detector to the level of single charges on the island.
This will allow to experimentally access the full electron counting statistics
to study the suppression of fluctuations in feedback controlled quantum
transport. Such an enhanced device would also enable us to address a key issue
in both metrology and information processing, namely the accuracy and
fidelity of the single charge source. An extension of our device
to a composite quantized current source composed of three pumps with two
islands in series will allow us to realize the feedback controlled ``perfect"
current source anticipated by Brandes:~\cite{Brandes2010} Due to the mesoscopic
charge feedback demonstrated in this paper, the current through the second and
third pump will be locked onto the first one. Simultaneously monitoring single
charge fluctuations on both islands will allow us to detect and distinguish
current fluctuations due to a single extra or missed electron.~\cite{Wulf2008}
Feeding back this error signal onto the gate voltage controlling the first pump
will then suppress the long-time current fluctuations of this composite
quantized current source, realizing a ``perfect" dc current source.

%\begin{acknowledgments}
We would like to thank T.\ Weimann for support of the clean room processing and C.\
Fricke for assistance in performing the measurements. We acknowledge financial
support by the DFG and by QUEST. This work has been supported by EURAMET
joint research project with European Community's 7th Framework
Programme, ERANET Plus under Grant Agreement
No. 217257.
%\end{acknowledgments}

%

\end{document}